\documentclass[journal,twoside,web]{ieeecolor}
\usepackage{generic}
\usepackage{cite}
\usepackage{amsmath,amssymb,amsfonts}
\usepackage{algorithmic}
\usepackage{graphicx}
\usepackage{textcomp}
\usepackage{array, booktabs, makecell}
\usepackage{multirow}
\usepackage{mathtools}

\def\Fig#1{Fig.~\ref{fig:#1}}

\def\Tab#1{Table~\ref{tab:#1}}
\def\Sec#1{Section~\ref{sec:#1}}

\def\BibTeX{{\rm B\kern-.05em{\sc i\kern-.025em b}\kern-.08em
    T\kern-.1667em\lower.7ex\hbox{E}\kern-.125emX}}
\begin{document}
\title{A Multimodal Perceived Stress Classification Framework using Wearable Physiological Sensors}
\author{Muhammad Majid, Aamir Arsalan, and Syed Muhammad Anwar
%\thanks{Manuscript received April 19, 2022; revised August 26, 2015.}
\thanks{M. Majid is with the Signal Image Multimedia Signal Processing and LEarning (SIMPLE) Research Group, Department of Computer Engineering, University of Engineering and Technology Taxila, Taxila, Pakistan (e-mail: m.majid@uettaxila.edu.pk.). A. Arsalan is with Department of Software Engineering, Fatima Jinnah Women University, Rawalpindi, Pakistan. S. M. Anwar is with Sheikh Zayed Institute for Pediatric Surgical Innovation, Children's National Hospital, Washington DC, USA.}
}

\maketitle

\begin{abstract}
Mental stress is a largely prevalent condition known to affect many people and could be a serious health concern. The quality of human life can be significantly improved if mental health is properly managed. Towards this, we propose a robust method for perceived stress classification, which is based on using multimodal data, acquired from forty subjects, including three (electroencephalography (EEG), galvanic skin response (GSR), and photoplethysmography (PPG)) physiological modalities. The data is acquired for three minutes duration in an open eyes condition. A perceived stress scale (PSS) questionnaire is used to record the stress of participants, which is then used to assign stress labels (two- and three classes). Time (four from GSR and PPG signals) and frequency (four from EEG signal) domain features are extracted. Among EEG-based features, using a frequency band selection algorithm for selecting the optimum EEG frequency subband, the theta band was selected. Further, a wrapper-based method is used for optimal feature selection. Human stress level classification is performed using three different classifiers, which are fed with a fusion of the selected set of features from three modalities. A significant accuracy (95\% for two classes, and 77.5\% for three classes) was achieved using the multilayer perceptron classifier.
\end{abstract}


\section{Introduction}
\label{sec:introduction}
\IEEEPARstart{O}{nly} a few decades ago, a major threat to the existence of human life included diseases such as pneumonia and tuberculosis~\cite{de2011stress}. However, in today's fast-paced life, people rarely die due to these diseases because of the availability of state-of-the-art treatment facilities. Instead, the mortality rate is found to be high in people suffering from ailments such as cardiovascular disease, cancer, and diabetes~\cite{world2020technical}. One of the most important and common causes of such diseases is stress~\cite{salleh2008life}. Homeostasis refers to the level of glucose flowing through blood vessels that are required to keep the body temperature to a balanced level. When a stressful situation occurs, the homeostatic balance of a person gets disturbed, and as a result, the equilibrium in the human body is effected~\cite{bakker2011s}. Stress can be originated due to internal as well as external factors resulting in the release of hormones like adrenaline and cortisol. These hormones flow through the body causing a change in different physiological parameters like the heart rate, brain activity, and blood flow~\cite{mcewen2008central}. In case, the stress condition persists for a longer duration, it could result in a weakening of the immune system causing a delay in recovery from various infections and ailments~\cite{segerstrom2004psychological}. In addition to the effect of stress on physical health, it also affects our performance and passion at work and behavior towards daily life activities. A study has reported that an increase in the stress level of employees in the workplace resulted in an overall degradation in performance~\cite{sharma2012objective}. In general, coping with stress is not only critical to personal health but also for a better and healthy society.

Human stress can be classified into acute (short-term or instantaneous) and chronic (long-term or perceived) stress~\cite{picard2016automating}. Acute stress is caused by the mental strain from events of the recent past and or those going to occur soon. This could include various forms of social interaction, for instance, an argument with a family member or an anxious feeling while meeting new people. The scenarios could be very different and personal, like bad economic conditions, an unhappy marriage, or a mentally demanding workload, and could all be a cause of chronic stress~\cite{england2012epilepsy}. A common thing in all these situations includes a sudden change in human circumstances and things happening differently from normal. Chronic stress needs to be addressed properly otherwise it can cause serious health problems~\cite{ryvlin2013incidence}. Muscular tension, back pain, digestion problems, cardiac arrhythmia, heart attack, or even sudden death are the consequences of acute stress~\cite{miller2016stress}. Effects of chronic stress on the health of an individual are similar to acute stress but can cause more damage to the physical conditions of a person. Some of the health problems resulting from chronic stress include hypertension~\cite{pickering2001mental}, irritable bowel syndrome~\cite{monnikes2001role}, depression, and anxiety disorders~\cite{herbert1997fortnightly}.

Keeping into consideration these serious consequences of stress (both acute and chronic) to human health and social well-being, it is of utmost importance to identify and cure stress conditions. Towards stress identification, subjective measures have been used which include filling out questionnaires designed by psychologists. One of the commonly used questionnaires to measure perceived stress is the perceived stress scale (PSS)~\cite{cohen1983global}. On the other hand, objective measures of stress are composed of both physical and physiological measures. Physical measures of stress show changes in the form of facial expressions~\cite{deschenes2015facial}, pupil dilation~\cite{wang2009pupil}, and eye blink rate~\cite{gowrisankaran2012asthenopia}. However, physiological measures of stress require sensors to be placed on the human body to measure physiological bio-markers like electroencephalography (EEG)~\cite{jebelli2018continuously}, galvanic skin response (GSR)~\cite{visnovcova2016complexity}, electrocardiography (ECG)~\cite{charbonnier2018multi}, and heart rate variability (HRV)~\cite{hernando2016inclusion}.

Multimodal schemes are beneficial, particularly for solving an ill-posed problem~\cite{lahat2014challenges}. In the context of human stress measurement, it has been found that the fusion of data from multiple modalities increased the accuracy of the human stress measurement system~\cite{can2020real}. Some researchers have taken advantage of this fact by combining different modalities for the improvement of human stress measurement frameworks~\cite{wijsman2013wearable,zhai2006stress}. An extensive review of the stress assessment schemes available in the literature using both wearable and non-wearable sensors is presented in \cite{arsalan2022mental}.

\section{Related Work}
\label{sec:rw}
In literature, the assessment of human stress has been widely studied using data from different physiological sensors. Most of these studies have focused on instantaneous stress which is induced by standard laboratory stressors for instance the mental arithmetic task~\cite{hassellund2010long}. However, the classification of perceived or chronic stress has only been explored in a limited number of studies. For instance, a study aimed at developing an association between the PSS score and EEG signal acquired from a subject was presented in~\cite{saeed2015psychological}. A dominant beta-band activity has been reported in individuals experiencing high perceived stress. Another EEG-based study for the quantification of human stress was presented in~\cite{saeed2017quantification}. The study established the fact that by using regression, the beta band of the EEG signal can successfully predict the PSS score of the participants with a confidence level of $94\%$. EEG signals have been used for perceived stress classification using a correlation-based feature subset selection method~\cite{saeed2018selection}. The gamma and beta band of the EEG signal was found to be strongly associated with the participant's PSS score. An association between the temporal characteristics of the EEG signal and the PSS score was established in~\cite{luijcks2015influence}. The study concluded that users with higher perceived stress scores have stronger activation in the delta and theta band of the EEG signal during the post-stimulus phase of the experiment. Another EEG-based study to identify the phase of experiment suitable for the classification of perceived stress using EEG signals was presented in~\cite{arsalan2019classification}. The authors proposed a feature selection algorithm and concluded that by using the pre-activity phase, the theta band of the EEG signals is optimum for the classification of perceived stress. Average classification accuracy of $92.85\%$ and $64.28\%$ was reported for binary and multi-class (three) classification problems. An EEG-based long-term stress measurement scheme using stress labels obtained from both the PSS questionnaire and interview with the psychologist was presented in \cite{saeed2020eeg}. An accuracy of 85.20\% for two-level stress classification was reported. Another EEG signal-based perceived stress score prediction scheme (regression-based) was able to achieve a root mean square error value of 2.36 \cite{gillani2021prediction}.

\begin{figure*}
\begin{center}
\begin{tabular}{c}
\includegraphics[width=180mm]{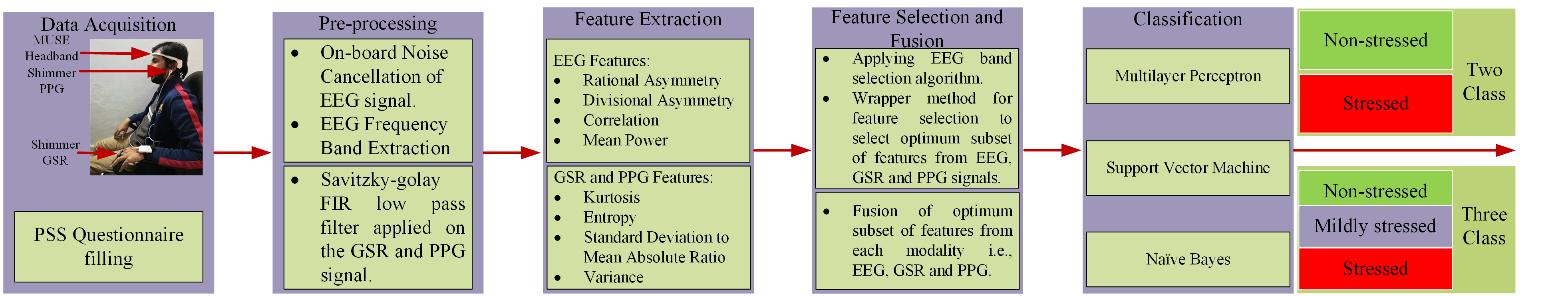}
\end{tabular}
\end{center}
\caption
{ \label{fig:fig1}
{Our proposed perceived stress level detection framework using wearable sensors.}}
\end{figure*}

A chronic stress measurement mechanism using GSR signals acquired during the state of sitting, sleeping, and standing was presented in~\cite{panigrahy2017study}. Stress and relaxed conditions were discriminated with a classification accuracy of 76.5\%. A human perceived stress measurement system using smartphone photoplethysmography and thermal imaging was presented in~\cite{cho2019instant} with a reported average classification accuracy of 78.3\%. A multimodal (including EEG, PPG, and GSR data) perceived stress classification method used only time-domain features~\cite{arsalan2019classification_EMBC}. An accuracy of $75\%$ was achieved for the two-class classification. Another hybrid approach of perceived stress classification using heart rate variability (HRV) and accelerometer sensor was presented in~\cite{wu2015modeling}. The study reported an average classification accuracy of 85.7\%. Another cluster-based approach for perceived stress measurement using a hybrid of physiological sensors of EEG, electrocardiography (ECG), electromyography (EMG), and GSR was presented in~\cite{xu2014cluster} with a reported classification accuracy of 85.2\%.

Most perceived stress assessment studies available in the literature have used a single modality and only two-class classification problem has been addressed except the study available in~\cite{arsalan2019classification}, which has also analyzed three-class problem. To the best of our knowledge, a study aimed at classifying the perceived stress into two- and three classes using a fusion of frequency- (from EEG) and time-domain (GSR and PPG) features has not been presented to date. In this study, the PSS questionnaire is used for labeling the participants into two- and three classes based on their scores followed by the acquisition of EEG, GSR, and PPG signals. We further performed feature selection for all three modalities and applied frequency band selection to EEG data. The selected features were then used to classify stress levels in to multiple (two and three) classes. Towards this, we have the following major contributions,
\begin{enumerate}
    \item We perform a multi-class perceived stress classification by fusing features extracted from data acquired using multiple physiological modalities (including EEG, GSR, and PPG).
    \item The statistical significance of various features, such as absolute power from different EEG subbands, skin resistance, and volumetric blood flow is analyzed for different stress levels. 
    \item A comprehensive analysis on selecting the appropriate features and frequency bands (for EEG data) is performed, resulting in a significant improvement in stress level classification at multiple levels.  
\end{enumerate}

The organization of the paper is as follows, \Sec{rw} presents the perceived human stress measurement schemes available in literature followed by the proposed methodology in \Sec{pm}. Results of our proposed stress classification scheme are presented in \Sec{simu}. A comprehensive comparison with existing models and discussion is presented in \Sec{disc}, which is followed by conclusions in \Sec{conc}.

\section{Our Proposed PSS Classification Protocol}\label{sec:pm}
The proposed multimodal framework for a multi-class perceived stress classification scheme (\Fig{fig1}) consists of four major blocks. This includes data acquisition, data pre-processing, feature extraction (in time and frequency domains) and selection, and stress level classification. Each of these steps is explained in detail in the following sub-sections.

\subsection{Data Acquisition}
\subsubsection{Participants}
Forty healthy participants, which included both males and females ($20$ males and $20$ females) with ages in the range $18$-$40$ years ($\mu$ = 24.85 and $\sigma$ = 6.69), participated in the experiment. Each participant had a minimum of twelve years of education and voluntarily participated in this study (consent forms were obtained). Moreover, non of the participant reported any mental disorder or physical disability. The experimental design for our proposed framework follows the protocols presented in the Helsinki declaration. This study was approved by the Board of Advanced Studies, Research, and Technological Development at the University of Engineering and Technology Taxila, Pakistan.
\subsubsection{Apparatus}
EEG data acquisition in this study was performed using the MUSE headband. MUSE is easy to wear and portable headband with dry electrodes. In particular, MUSE consists of two frontal (at positions $AF7$ and $AF8$), two temporal (at positions $TP9$ and $TP10$), and a reference (at $Fpz$) electrode. The sampling rate of the MUSE headband is $256$ Hz. Further, MUSE is connected to the Muse Monitor application using Bluetooth protocol. The recorded EEG signals are saved in the form of a comma-separated value (CSV) file, which is transmitted to a computer for further analysis and processing.

PPG and GSR (sensors placed on the left-hand fingers and left ear lobe) signal acquisition in this study was performed using the Shimmer GSR+ module. This module is reliable and easy-to-use equipment for recording PPG and GSR signals and has been widely used in different research studies~\cite{gonzalez2019physiological}. Shimmer GSR+ module performs data acquisition at a sampling rate of $256$ Hz, using the ShimmerCapture desktop application which streams data via a Bluetooth connection. The recorded data is saved in a CSV file.

\subsubsection{Experimental Procedure}
Each participant in the experiment was guided to the data acquisition room and was given a comfortable sitting place. The experimental procedure was thoroughly explained to the participant and the demographic details and written consent to participate in the experiment were obtained. Next, the subject was given the PSS questionnaire to be filled out. PSS is a questionnaire consisting of $10$ items used to measure the perceived stress (over the last $30$ days) of an individual. In particular, the subject can answer each question in the questionnaire from a minimum value of $0$ to a maximum value of $4$. A response of $0$ means that a certain event rarely happened and response of $4$ means that a certain event occurred very frequently in the last month. The final score for the participant is the sum of the scores of the individual's items on the questionnaire. The PSS score of the participant can range on a scale from $0$ to $40$ and be used for labeling the participants for both two- and three classes.

After filling out the PSS questionnaire, the MUSE headband was placed on the scalp of the subject and GSR and PPG sensors were placed on the fingers of the left hand and the left ear lobe of the participant, respectively. The participants were seated comfortably in a chair and were instructed to keep their eyes open while focusing on a blank screen. EEG, GSR, and PPG data acquisition were performed in this state for a duration of $3$ minutes, after which the participant was free to leave the experiment room.

\subsection{Pre-processing}
The physiological signals were pre-processed for improving the signal quality before feature extraction and selection. EEG signals were processed using the MUSE on-board noise cancellation. The EEG signal was deemed to be clean if the value of statistical properties (such as variance and kurtosis) was less than a certain threshold. Fast Fourier transform (with a window size of 256 and an overlap of 90\%) of the raw EEG signals using the on-board signal processing module of the MUSE headband was calculated to obtain the EEG frequency bands. The resulting EEG frequency bands include delta (0–4 Hz), theta (4–7 Hz), alpha (8–12 Hz), beta (12–30 Hz), and gamma (30–50 Hz) bands.

The PPG and GSR signals were passed through a Savitzky-Golay filter for data smoothing. Savitzky-Golay is a finite impulse response (FIR) low pass filter based on the least square polynomial approximation data smoothing technique~\cite{savitzky1964smoothing}. In this study, a third-order polynomial was used for the smoothing of the physiological signals.

\subsection{Feature Extraction}
After pre-processing, EEG, GSR, and PPG signals were subjected to feature extraction. In particular, features were extracted in both the time and frequency domains. Features extracted from the EEG signal include differential asymmetry (DASM), rational asymmetry (RASM), correlation (C), and mean power ($P_{mean}$). For GSR and PPG signals, features included kurtosis (K), entropy (E), the standard deviation to mean absolute ratio (SdMar), and variance ($\sigma^2$). The description and the feature vector length (FVL) for each of the extracted features are elaborated below.

\subsubsection{EEG Features}
DASM is computed by the subtraction of the absolute powers of the hemispheric asymmetry electrodes. A feature vector length of 10, i.e., five from each of the two asymmetric channel pairs was obtained. RASM is calculated as a ratio of the absolute powers of the hemispheric asymmetry electrodes. Similar to DASM, an FVL of 10, i.e., five from each of the two asymmetric channel pairs were obtained. Correlation is the measure of the amount of variation of one value with another. Correlation between the hemispheric asymmetry electrodes was used as a feature in our study. An FVL of 10, i.e., five from each of the two asymmetric channel pairs was obtained for correlation. The average absolute power for each frequency band (four) from each EEG channel of MUSE resulted in an FVL of 20 i.e., 5 from each of the MUSE headband EEG electrodes.

\subsubsection{GSR and PPG Features}
Kurtosis is a statistical property explaining the distribution of data and is computed as a fraction of the fourth standardized moment and the square of the second standardized moment i.e., the variance. Entropy computes the amount of uncertainty or randomness in a given data sample. Standard deviation to mean absolute ratio is calculated as a ratio of standard deviation and mean for the GSR and PPG data. Variance is measured as an amount of spread among the value in a given data sample, which in our case is the GSR and PPG data. An FVL of 2 (i.e., 1 for GSR and 1 for PPG data) from each of these four features (kurtosis, entropy, standard deviation to mean absolute ratio, and variance) resulting in a total of 8 features were obtained.

\subsection{Feature Selection and Fusion}
The recorded EEG data were processed using the band selection algorithm, which has been developed particularly for human stress measurement in~\cite{arsalan2019classification}. The objective is to identify those EEG frequency bands which yield the highest stress level classification accuracy. The aforementioned band selection algorithm works by selecting those EEG frequency bands which maximize the classification accuracy from among all the frequency band combinations. To select the optimum band, 1000 iterations were performed for each combination. Moreover, the features extracted from the selected EEG frequency band and the features from the GSR and the PPG signal are subjected to the wrapper method for feature selection to select the subset of features that are highly correlated to the class labels and least correlated to each other. The wrapper method for feature selection is a classifier-dependent technique and selects a different feature subset for each classifier which yields the highest classification accuracy. The final feature vector was obtained by a fusion of a selected subset of features. Feature fusion can be applied in two variations which include early feature fusion (EFF) and late feature fusion (LFF). In EFF, extracted features from different modalities are concatenated, and then feature selection is applied to the concatenated feature vector to obtain an optimum subset of features. Whereas in LFF, feature selection is applied to features extracted from various modalities. These selected features are concatenated to obtain an optimum feature subset. In this study, we have used the LFF scheme, and the obtained feature vector was subjected to classification.

\subsection{Classification}
Perceived stress classification into two- and three classes was performed using three different classifiers which include a support vector machine, a multilayer perceptron, and the Naive Bayes algorithm. The details of each of the algorithm is given as follows.

\subsubsection{Support Vector Machine (SVM)}
SVM is a supervised machine learning algorithm that can be used for classification as well as regression problems. SVM algorithm works by estimating a hyper-plane in the N-dimensional space to classify the data points distinctly. The objective is to find an optimum hyperplane that has the maximum margin i.e., there should be a maximum distance between the data points of the two classes or the support vectors. Future data points can be classified with better confidence by maximizing the margin between the data points. In this study, SVM with radial basis function (RBF) kernel was used with $\gamma$ = 0.01 and C = 10.

\subsubsection{Multilayer Perceptron (MLP)}
MLP is a type of feed-forward artificial neural network. The simplest kind of MLP consists of a three-layer network i.e., an input layer, a hidden layer, and an output layer.  Each neural network model consists of input and output layers and the number of hidden layers varies according to the complexity of the problem at hand. The neural network architecture used in this study had four hidden layers. The learning rate hyper-parameter was set to 0.3, and the momentum value of 0.2 was selected after experimentation. 

\subsubsection{The Naive Bayes (NB)}
The Naive Bayes classifier is a probabilistic supervised machine learning model, which is based on the Bayes theorem. The Naive Bayes algorithm is based on the assumption that each of the extracted features contributes equally to the outcome and all the features are independent of each other. The algorithm can be used for binary as well as multi-class classification problems.

\section{Results}\label{sec:simu}
In this section, statistical analysis as well as perceived stress classification results in two (non-stressed and stressed) and three (non-stressed, mildly stressed, and stressed) classes using EEG, GSR, and PPG signals are reported.

\subsection{Data Labeling}
The labeling of the subjects into two and three classes was performed using the PSS questionnaire score. The mean and the variance of the PSS score of all the participants was $\mu$ = 22 and $\sigma$ = 7.15. For the two-class problem, the participants with a scoreless than and equal to the mean value are labeled as non-stressed, whereas the participants with a score greater than the mean value are labeled as stressed participants. For three-class classification, labeling of the participants are performed in such a manner that participants with PSS score ranging from $0$ to $\lceil{\mu -\frac{\sigma}{2}}\rceil$ i.e., ($0$ - $18$) are labeled as non-stressed, participants having PSS score ranging from $\lceil{\mu -\frac{\sigma}{2}}\rceil+1$ to $\lceil{\mu +\frac{\sigma}{2}}\rceil-1$ i.e., ($19$ - $26$) are labeled as mildly stressed, and the participants having PSS score greater than $\lceil{\mu +\frac{\sigma}{2}}\rceil$ to $40$ i.e., ($27$ - $40$) are labeled as stressed. Based on this labeling, $22$ participants are labeled as non-stressed and $18$ participants are labeled as stressed participants for the two-class problem. For three class problem, $12$ participants are labeled as non-stressed, $19$ participants are labeled as mildly stressed and $9$ participants are labeled as stressed participants.

\subsection{Statistical Analysis}
A t-test was used to compute statistically significant differences between the two classes i.e., stressed and non-stressed. While for three classes i.e., non-stressed, mildly stressed, and stressed, \textbf{AN}alysis \textbf{O}f \textbf{VA}riance (ANOVA) was applied to different groups of the recorded physiological signals. T-test and ANOVA were applied to the absolute power of each band of the EEG signal, skin resistance (from the GSR signal), and blood volumetric flow (from the PPG signals). For EEG signals, applying the t-test on two-class problem showed that theta band for electrodes at position $TP9$ (p-value: 0.039) and $TP10$ (p-value: 0.041) was significant. Beta band from electrode $TP9$ had a p-value of $0.049$. Moreover, for three-class problem, applying ANOVA showed that theta band for electrodes at $TP9$ (p-value: $0.015$) and $TP10$ (p-value: $0.034$), and beta band for electrode at $TP9$ (p-value: $0.0245$) were statistically significant. Whereas, none of the frequency bands of any other channel of the MUSE headband was found to be statistically significant with a p-value less than $0.05$. Moreover, for GSR data there exist a partial significance for the two-class problem (p-value: $0.058$), whereas for the three-class problem there exists a strong statistical significance (p-value: $0.035$). For the PPG signal, there exists statistical significance for two (p-value: $0.025$) and three classes (p-value: $0.049$).

\begin{figure}
\begin{center}
\begin{tabular}{c}
\includegraphics[width=90 mm]{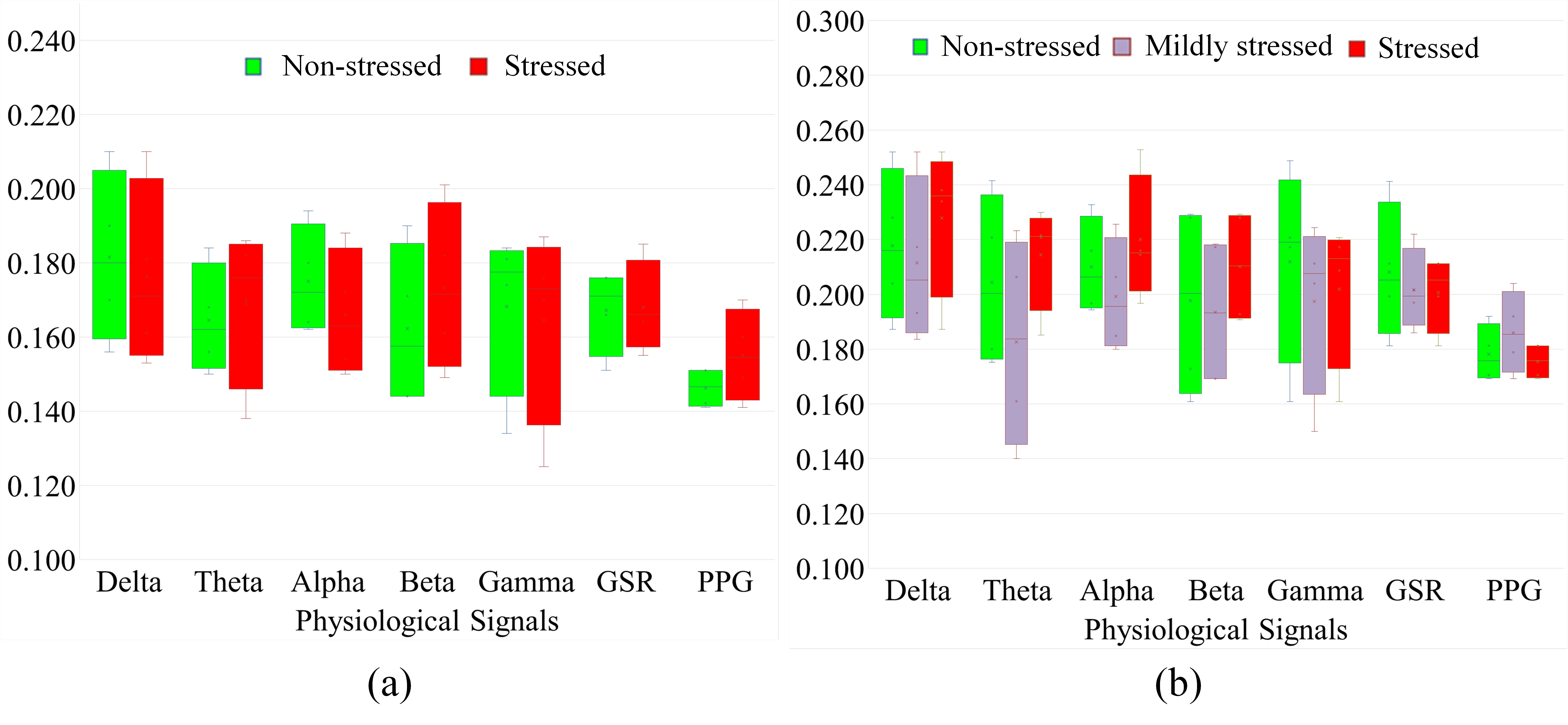}
\end{tabular}
\end{center}
\caption
{ \label{fig:fig2}
{Boxplot for all the frequency bands for the EEG, GSR and PPG signals for (a) two class problem i.e., non-stressed and stressed groups (b) three class problem i.e., non-stressed, mildly stressed and stressed participants.}}
\end{figure}

\begin{figure}
\begin{center}
\begin{tabular}{c}
\includegraphics[width=90 mm]{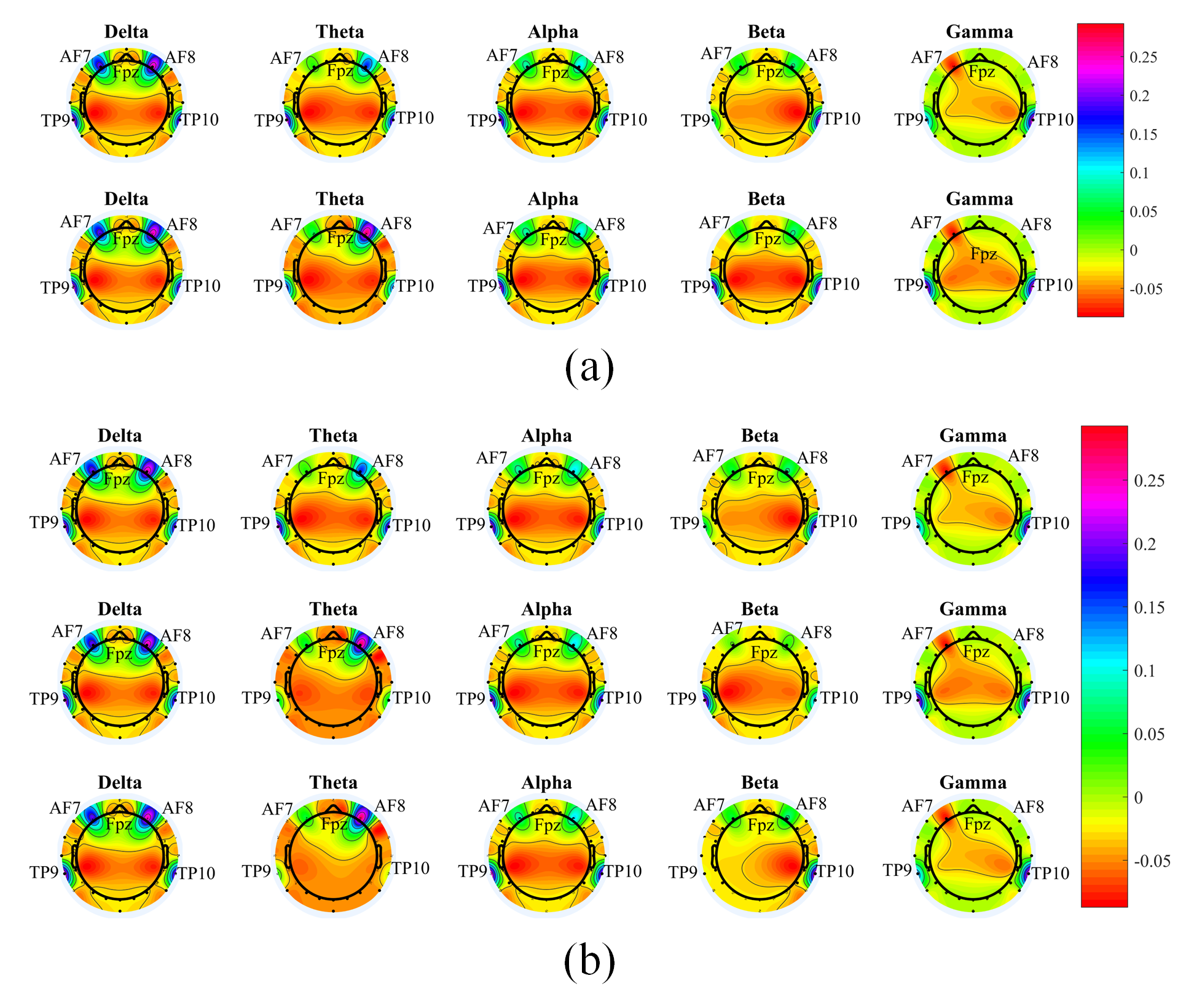}
\end{tabular}
\end{center}
\caption
{ \label{fig:fig3}
{Brain visualization using the average power spectral density in (a) two class problem i.e., non-stressed and stressed groups (b) three class problem i.e., non-stressed, mildly stressed and stressed participants.}}
\end{figure}

We further analyzed the results using box plots for two- and three-class classification as shown in \Fig{fig2}. For EEG signals, it is observed that for both two (\Fig{fig2} (a)) and three (\Fig{fig2} (b)) classes, theta band of the temporal channel $TP9$ and $TP10$ and the beta band of the channel $TP9$ are distinct from each other. For the GSR and PPG signals, there also exists a distinction in the box plots among non-stressed and stressed groups for the two-class problem (\Fig{fig2} (a)) and non-stressed, mildly stressed, and stressed groups for the three-class problem (\Fig{fig2} (b)) as supported by the p-values obtained from the t-test and ANOVA. These results are further validated using the brain visualizations, which represent the average EEG power spectral density (\Fig{fig3}). In these brain maps, the red color indicates a strong brain activity, whereas the orange color represents a weak brain activity. It is evident from these results that the theta band of channels $TP9$ and $TP10$ and beta band of $TP9$ has significantly different activities for two- as well as three classes.

\subsection{Feature Selection}
First, the relevant EEG signal bands were selected using a band selection algorithm~\cite{arsalan2019classification}. This algorithm is specifically designed to select the optimum frequency band for human stress classification whose features can be used for feature selection and classification purposes. In the current study, the band selection algorithm applied to the EEG signals selects the theta band as the optimum band because it yields the highest classification accuracy among all frequency bands. Next, the features from the selected frequency bands, GSR, and the PPG signals are subjected to the wrapper method for feature selection. The features selected for EEG signal include PS ($\theta_{TP9}$, $\theta_{TP10}$, $\theta_{AF8}$), C ($\theta_{TP9,TP10}$), RASM ($\theta_{TP9,TP10}$), DASM ($\theta_{AF7,AF8}$). Feature selected from the GSR signal includes standard deviation to mean absolute ratio and variance and features selected from the PPG signal include standard deviation to mean absolute ratio, entropy, and variance. These selected features were fused and further used to classify the perceived stress.

\begin{figure*}
\begin{center}
\begin{tabular}{c}
\includegraphics[width=180 mm]{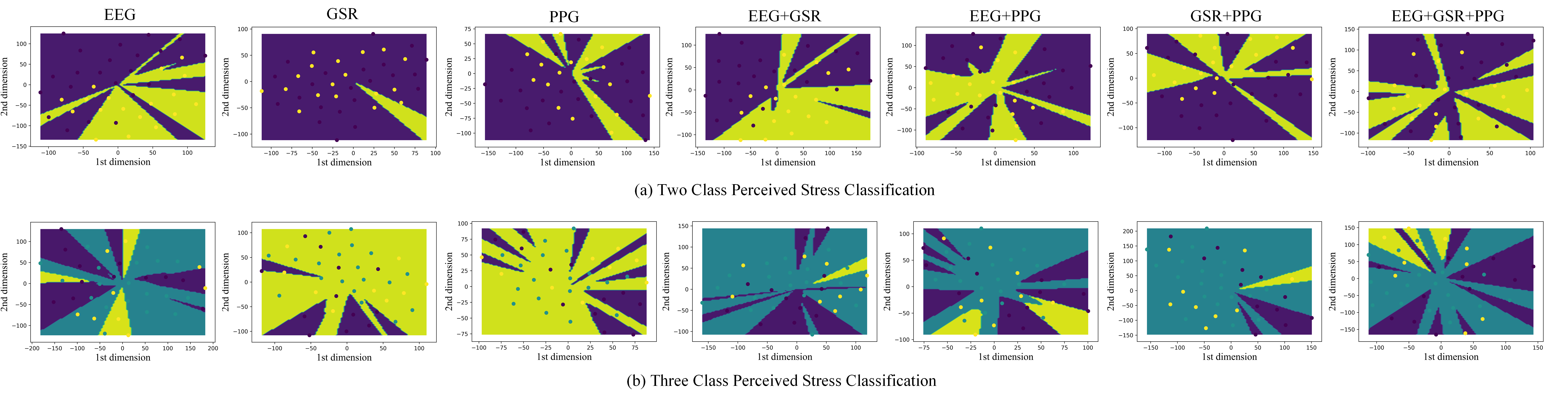}
\end{tabular}
\end{center}
\caption
{ \label{fig:fig4}
{Selected features obtained from different modalities represented in two dimensions using t-SNE with decision boundaries for First Row: Two class Second Row: Three class perceived stress classification.}}
\end{figure*}

Features from each modality along with all possible feature fusions are visualized using t-Distributed Stochastic Neighbor Embedding (t-SNE)~\cite{maaten2008visualizing} and Voronoi-based visualization schemes~\cite{migut2015visualizing}. t-SNE is a dimensionality reduction scheme widely used for visualizing the high dimensional data and the Voronoi-based visualization scheme is a mechanism to draw the decision boundaries by partitioning the plane based on a given set of points. \Fig{fig4} shows the feature visualization of the selected features of each modality as well as a fusion of these selected features in all possible combinations. It is evident from these visualizations that the lowest number of misclassified instances are obtained in the case of the fusion of selected features from all three signal sources.

\subsection{Classification Performance}
Classification of perceived stress in two- and three-classes was performed using three different classifiers i.e., MLP, SVM, and NB. The training and testing part of the classification algorithms was performed in Weka tool. The classifier performance was evaluated using a leave-one-out cross-validation scheme. For evaluation metrics accuracy, f-score, and kappa statistics were used (\Tab{tab2}). Various modalities were combined in different combinations (all possible combinations used) to come up with this comparison. The best performance was achieved when using MLP classifier and a fusion of features from all three modalities. In particular, an accuracy of $95\%$ (two classes) and $77.5\%$ (three classes) with a feature vector length of $11$ was achieved. Moreover, a f-score of $0.94$ (two-classes) and $0.75$ (three classes), and a kappa statistics value of $0.84$ (two-classes) and $0.46$ (three classes) was achieved.

\begin{table}[t]
%\small
\caption{A summary of the classification performance (for two and three classes) evaluated using accuracy, F-measure and Kappa and considering various combinations of the modalities and classifiers used in this study. It can be seen (values in bold) that multimodal data consistently performs better in all evaluation metrics. $Acc$: Accuracy, $F_m$: f-score, $K$: kappa.}
\begin{center}
\scalebox{0.9}{
%\small\addtolength{\tabcolsep}{-2pt}
\begin{tabular}{|c|c|c|c|c|c|}
\hline
\thead{Modalities} & Classes & \thead{Classifier} &  \thead{$Acc$ (\%)} & \thead{$F_m$} & \thead{$K$} \\ \hline

\multirow{6}{*}{\thead{$EEG$}} & \multirow{3}{*}{Two} & MLP & 85.00 & 0.84 & 0.75 \\ \cline{3-6}
    & & SVM & 67.50 & 0.65 & 0.40 \\ \cline{3-6}
    & & NB  & 72.50 & 0.70 & 0.38 \\ \cline{2-6}

    & \multirow{3}{*}{Three} & MLP & 67.50 & 0.65 & 0.34 \\ \cline{3-6}
    & & SVM & 57.50 & 0.57 & 0.15 \\ \cline{3-6}
    & & NB  & 60.00 & 0.59 & 0.32 \\ \cline{1-6}

\multirow{6}{*}{\thead{$GSR$}} & \multirow{3}{*}{Two} & MLP & 50.00 & 0.49 & 0.24 \\ \cline{3-6}
    & & SVM & 37.50 & 0.37 & -0.24 \\ \cline{3-6}
    & & NB  & 42.50 & 0.41 & -0.19 \\ \cline{2-6}

    & \multirow{3}{*}{Three} & MLP & 47.50 & 0.46 & 0.18 \\ \cline{3-6}
    & & SVM & 40.00 & 0.39 & -0.10 \\ \cline{3-6}
    & & NB  & 45.00 & 0.44 & -0.14 \\ \cline{1-6}

\multirow{6}{*}{\thead{$PPG$}} & \multirow{3}{*}{Two} & MLP & 65.00 &0.64 & 0.43 \\ \cline{3-6}
    & & SVM & 57.50 & 0.57 & 0.14 \\ \cline{3-6}
    & & NB  & 72.50 & 0.70 & 0.37 \\ \cline{2-6}

    & \multirow{3}{*}{Three} & MLP & 52.50 & 0.51 & 0.36 \\ \cline{3-6}
    & & SVM & 47.50 & 0.46 & -0.06 \\ \cline{3-6}
    & & NB  & 42.50 & 0.41 & -0.02 \\ \cline{1-6}

\multirow{6}{*}{\thead{$EEG+GSR$}} & \multirow{3}{*}{Two} & MLP & 77.50 & 0.74 & 0.46 \\ \cline{3-6}
    & & SVM & 57.50 & 0.53 & 0.06 \\ \cline{3-6}
    & & NB  & 62.50 & 0.59 & 0.18 \\ \cline{2-6}

    & \multirow{3}{*}{Three} & MLP & 65.00 & 0.62 & 0.23 \\ \cline{3-6}
    & & SVM & 57.50 & 0.56 & 0.15 \\ \cline{3-6}
    & & NB  & 52.50 & 0.49 & 0.19 \\ \cline{1-6}    
 
\multirow{6}{*}{\thead{$EEG+PPG$}} & \multirow{3}{*}{Two} & MLP & 80.00 & 0.81 & 0.61 \\ \cline{3-6}
    & & SVM & 72.50 & 0.66 & 0.35 \\ \cline{3-6}
    & & NB  & 70.00 & 0.65 & 0.27 \\ \cline{2-6}

    & \multirow{3}{*}{Three} & MLP & 67.50 & 0.63 & 0.28 \\ \cline{3-6}
    & & SVM & 62.50 & 0.60 & 0.19 \\ \cline{3-6}
    & & NB  & 62.50 & 0.59 & 0.20 \\ \cline{1-6}     

\multirow{6}{*}{\thead{$GSR+PPG$}} & \multirow{3}{*}{Two} & MLP & 75.00 & 0.72 & 0.44 \\ \cline{3-6}
    & & SVM & 72.50 & 0.68 & 0.37 \\ \cline{3-6}
    & & NB  & 67.50 & 0.63 & 0.28 \\ \cline{2-6}

    & \multirow{3}{*}{Three} & MLP & 57.50 & 0.57 & 0.14 \\ \cline{3-6}
    & & SVM & 52.50 & 0.51 & 0.07 \\ \cline{3-6}
    & & NB  & 47.50 & 0.45 & -0.04 \\ \cline{1-6}

\multirow{6}{*}{\thead{$EEG+GSR+PPG$}} & \multirow{3}{*}{Two} & \textbf{MLP} & \textbf{95.00} & \textbf{0.94} & \textbf{0.84} \\ \cline{3-6}
    & & SVM & 77.50 & 0.76 & 0.48 \\ \cline{3-6}
    & & NB  & 82.50 & 0.81 & 0.57 \\ \cline{2-6}

    & \multirow{3}{*}{Three} & \textbf{MLP} & \textbf{77.50} & \textbf{0.75} & \textbf{0.46} \\ \cline{3-6}
    & & SVM & 65.00 & 0.63 & 0.29 \\ \cline{3-6}
    & & NB  & 67.50 & 0.66 & 0.28 \\ \cline{1-6}
\end{tabular}
}
\label{tab:tab2}
\end{center}
\end{table}

\begin{table}
\caption{The two class perceived stress classification confusion matrix when using MLP classifier.}
\begin{center}
\scalebox{1.0}{
\begin{tabular}{c|c|c|c|c}
\hline\hline
\thead{Stressed \\(S)} & \thead{Non-stressed \\(NS)} & Classified as & Precision & Recall \\ \hline
21   &  1  & NS  & 0.95  &  0.95 \\
1    &  17 & S  &  0.94 &  0.94 \\ \hline\hline
\end{tabular}
}
\label{tab:tab3}
\end{center}
\end{table}

\begin{table}
\caption{The three class perceived stress classification confusion matrix when using MLP classifier.}
\begin{center}
\scalebox{1.0}{
\begin{tabular}{c|c|c|c|c|c}
\hline\hline
NS & MS & S & Classified as & Precision & Recall \\ \hline
10 & 1  & 1 & NS=Non-stressed  & 0.76  &  0.83 \\
2  & 15 & 2 & MS=Mildly Stressed  &  0.83 &  0.78 \\
1  & 2  & 6 & S=Stressed & 0.66 & 0.66 \\ \hline\hline
\end{tabular}
}
\label{tab:tab4}
\end{center}
\end{table}

Confusion matrices along with the precision and recall values are presented in \Tab{tab3} and \Tab{tab4}, respectively. For the two-class problem, $21$ out of $22$ non-stressed subjects and $17$ out of $18$ stressed subjects were correctly classified. Moreover, for the three-class problem, $10$ out of $12$ non-stressed subjects were correctly classified, $15$ out of $19$ mildly stressed participants were correctly classified, and $6$ out of $9$ stressed participants were correctly classified. In terms of precision and recall, the MLP classifier has better performance when compared to SVM and NB classifiers.

\section{Discussion}\label{sec:disc}
In this study, a multimodal perceived stress classification framework is proposed. The comparison of the proposed scheme with the state-of-the-art methods available in the literature is presented in \Tab{tab5}. We were able to collect data from the largest number of participants (40) when compared to reported studies. The studies involving only EEG signals can be divided into single- and four-channel-based studies. The maximum accuracy achieved for single-channel EEG-based studies is $78.57\%$ for two-class classification, whereas for the four-channel EEG-based studies a maximum accuracy of $92.85\%$ and $64.28\%$ is achieved for two- and three-classes, respectively.

Apart from only EEG based studies,~\cite{cho2019instant} achieves an accuracy of $78.30\%$ using PPG and thermography modality,~\cite{wu2015modeling} achieves an accuracy of $85.7\%$ using HRV and accelerometer sensors,~\cite{panigrahy2017study} achieves an accuracy of $76.50\%$ using only GSR sensors, and~\cite{arsalan2019classification_EMBC} achieves an accuracy of $75\%$ using EEG, GSR, and PPG sensors. Moreover, the classification accuracy is reported for two classes. In comparison, our proposed scheme not only performs better but also considers a more challenging three-level classification. Some of the limitations of the proposed stress classification framework include a lack of testing of the proposed scheme in an out-of-lab environment and the non-availability enough data for applying deep learning-based methods to evaluate further improvement in the classification accuracy of the system.

\begin{table}
%\small
\caption{A comparative performance analysis of our proposed stress level classification scheme with state-of-the-art methods in terms of participants, modalities, classification schemes and accuracy.}
\begin{center}
%\scalebox{1}{
%\small\addtolength{\tabcolsep}{-2pt}
\begin{tabular}{|c|c|c|c|c|}
\hline
\thead{Method\\Year} &  \thead{Participants} &  Data Source & \thead{Classification \\Scheme} & \thead{Accuracy \\ (Classes)} \\ \hline
\thead{~\cite{arsalan2019classification}, \\ 2019} & 28 & EEG & MLP  & \thead{92.85\% \\ (2) \\ 64.28\% \\ (3)}\\ \hline
\thead{~\cite{arsalan2019classification_EMBC}, \\ 2019} & 28 & \thead{EEG, \\GSR, PPG} & \thead{MLP} & \thead{75.00\% \\ (2)} \\ \hline
\thead{~\cite{cho2019instant}, \\ 2019} & 17 & \thead{PPG,\\thermography} & \thead{Artificial \\Neural \\Network} & \thead{78.30\% \\ (2)} \\ \hline
\thead{~\cite{saeed2018selection}, \\ 2018} & 28 & EEG &  SVM  & \thead{78.57\% \\ (2)} \\ \hline
\thead{~\cite{saeed2017quantification}, \\ 2017} & 28 & EEG & \thead{Naive \\ Bayes} & \thead{71.40\% \\ (2)} \\ \hline
\thead{~\cite{panigrahy2017study}, \\ 2017} & 10 & GSR & J48 & \thead{76.50\% \\ (2)} \\ \hline
\thead{~\cite{saeed2015psychological}, \\ 2015} & 28 & EEG & SVM & \thead{71.42\% \\ (2)} \\  \hline
\thead{~\cite{wu2015modeling}, \\ 2015} & 8 & \thead{HRV, \\accelerometer} & \thead{Bagging \\ classifier} & \thead{85.70\% \\ (2)} \\ \hline
\textbf{\thead{Proposed, \\ 2022}} & \textbf{40} & \textbf{\thead{EEG,\\GSR, PPG}} & \textbf{MLP} & \textbf{\thead{95.00\% \\ (2) \\ 77.50\% \\ (3)}} \\ \hline

\end{tabular}
%}
\label{tab:tab5}
\end{center}
\end{table}

\section{Conclusion}\label{sec:conc}
In this paper, a perceived human stress classification framework using physiological signals from multiple modalities is presented. Four feature groups, which include rational asymmetry, divisional asymmetry, correlation, and power spectrum were extracted from the frequency bands of the EEG signal. Whereas for GSR and PPG signals, features included kurtosis, entropy, the standard deviation to mean absolute ratio, and variance. A band selection algorithm was applied to the features extracted from the EEG signals and the frequency band (theta band) yielding the highest classification accuracy was identified. Further, the wrapper method for feature selection was applied to these features. The human stress level was classified (in two- and three classes) using three different classifiers (SVM, NB, and MLP) and a fusion of features identified during the feature selection process. An accuracy of $95\%$ (two classes) and $77.50\%$ (three classes) was achieved using the MLP classifier. Future work for the proposed stress classification scheme can include a hybrid approach for labeling the data i.e., by using questionnaire scores and evaluation by a psychologist to further improve the system accuracy. Evaluation of the system performance in an out-of-lab environment and acquisition of data from more subjects can be performed to apply a convolutional neural network and other deep learning methods for stress classification.

\bibliographystyle{IEEEtran}
\bibliography{mybibfile}

\end{document}